# Historical and Philosophical Aspects of the Einstein World

Cormac O'Raifeartaigh

*School of Science and Computing, Waterford Institute of Technology, Cork Road, Waterford, Ireland*
*Email: coraifeartaigh@wit.ie*

This article presents a brief review of some historical and philosophical aspects of Einstein's 1917 paper *'Cosmological Considerations in the General Theory of Relativity'*, a landmark work that denoted the starting point of modern theoretical cosmology. Our presentation includes a discussion of Einstein's early views of issues such as the relativity of inertia, the curvature of space and the cosmological constant. Particular attention is paid to lesser-known aspects of Einstein's paper such as his failure to test his model against observation, his failure to consider the stability of the model and a slight mathematical confusion concerning the introduction of the cosmological constant term. Taken in conjunction with his later cosmological works, we find that Einstein's approach to cosmology was characterized by a pragmatic search for the simplest model of the universe that was consistent with the principles of relativity and with contemporaneous astronomical observation.

1. **Introduction**

There is little doubt that Einstein's 1917 paper *'Cosmological Considerations in the General Theory of Relativity'* (Einstein 1917a) constituted a key milestone in 20$^{th}$ century physics. The paper introduced the first relativistic model of the universe, sometimes known as 'Einstein's Static Universe' or the 'Einstein World' and marked the starting point of modern theoretical cosmology.
To be sure, a description of the basic physics of the Einstein World can be found in any standard textbook on modern cosmology (Harrison 2000 pp 355-357; Coles and Lucchin 2002 pp 26-28). However, while many accounts have been written of the development of theoretical cosmology from this point onwards, there have been surprisingly few detailed historical analyses of Einstein's 1917 paper itself.[1] This article presents a brief synopsis of our recent centenary review of the paper (O'Raifeartaigh et al. 2017), with an emphasis on lesser-known aspects of the work such as Einstein's failure to test his model against observation, his failure to consider the stability of the model and a slight mathematical confusion concerning the introduction of the cosmological constant term. We also consider Einstein's underlying approach to cosmology in the light of his later cosmological works.

## 2. Historical context of the Einstein World

*(i) Biographical context*

Einstein's manuscript *'Kosmologische Betrachtungen zur allgemeinen Relativitätstheorie'* or *'Cosmological Considerations in the General Theory of Relativity'* (Einstein 1917a) was read to the Prussian Academy of Sciences on February 8th 1917 and published by the Academy on February 15th of that year. Thus the paper, a sizeable ten-page memoir that was to play a seminal role in 20$^{th}$ century cosmology, appeared only eleven months after the completion of Einstein's greatest and most substantial work, *'Die Grundlage der allgemeinen Relativitätstheorie'* or *'The Foundations of the General Theory of Relativity'* (Einstein 1916a).[2] The short interval between these two monumental papers is astonishing given that Einstein completed many other works during this period and that he suffered a breakdown in health in early 1917.[3]

On the other hand, it is no surprise from a scientific point of view that Einstein's first foray into cosmology should occur so soon after the completion of the general theory of relativity. After all, it was a fundamental tenet of the general theory that the geometric structure of a region of space-time is not an independent, self-determined entity, but determined by mass-energy (Einstein 1916a). Thus, considerations of the universe at large posed an important test for the new theory. As Einstein later remarked to the Dutch astronomer Willem de Sitter:*"For me, though, it was a burning question whether the relativity concept can be followed through to the finish, or whether it leads to contradictions. I am satisfied now that I was able to think the idea through to completion without encountering contradictions"* (Einstein 1917b). Indeed, it is clear from Einstein's correspondence of 1916 and early 1917 that cosmic considerations – in the sense of the problem of boundary conditions at infinity – were a major preoccupation in the immediate aftermath of the discovery of the covariant field equations (Schulmann et al. 1998 pp 352-355).

---

[1] Some exceptions are (Kerzberg 1989; Realdi and Peruzzi 2009; Smeenk 2014).
[2] The 'Grundlage' paper was submitted to the *Annalen der Physik* on March 20$^{th}$ 1916 and appeared in print on May 11$^{th}$ of that year.
[3] These works included technical papers on quantum theory, gravitational waves, general relativity and a popular book on relativity (O'Raifeartaigh et al. 2017).

*(ii) Cosmology before 1917*

Few quantitative models of the universe were proposed before 1917. One reason was the existence of several puzzles associated with the application of Newton's universal law of gravity to the universe as a whole. For example, it was not clear how a finite Newtonian universe would escape gravitational collapse, as first pointed out by the theologian Richard Bentley, a contemporary of Isaac Newton. Newton's response was to postulate a universe infinite in spatial extent in which the gravitational pull of the stars was cancelled by opposite attractions. However, he was unable to provide a satisfactory answer to Bentley's observation that such an equilibrium would be unstable.[4]

Pioneering work on non-Euclidean geometries in the late 19th century led some theoreticians to consider the possibility of a universe of non-Euclidean geometry. For example, Nikolai Lobachevsky considered the case of a universe of hyperbolic (negative) spatial curvature and noted that the lack of astronomical observations of stellar parallax set a minimum value of 4.5 light-years for the radius of curvature of such a universe (Lobachevsky 2010). On the other hand, Carl Friedrich Zöllner noted that a cosmos of spherical curvature might offer a solution to Olbers' paradox[5] and even suggested that the laws of nature might be derived from the dynamical properties of curved space (Zöllner 1872). In the United States, astronomers such as Simon Newcomb and Charles Sanders Peirce took an interest in the concept of a universe of non-Euclidean geometry (Newcomb 1906; Peirce 1891 pp 174-175), while in Ireland, the astronomer Robert Stawall Ball initiated a program of observations of stellar parallax with the aim of determining the curvature of space (Ball 1881 pp 92-93; Kragh 2012a). An intriguing theoretical study of universes of non-Euclidean geometry was provided in this period by the German astronomer and theoretician Karl Schwarzschild, who calculated that astronomical observations set a lower bound of 60 and 1500 light-years for the radius of a cosmos of spherical and elliptical geometry respectively (Schwarzschild 1900). This model was developed further by the German astronomer Paul Harzer, who considered the distribution of stars and the absorption of starlight in a universe of closed geometry (Harzer 1908 pp 266-

---

[4] See (Norton 1999; Kragh 2007 pp 72-74) for a discussion of the Newton-Bentley debate.
[5] This well-known problem concerned the difficulty of reconciling the darkness of the night sky with a universe infinite in space and time (Kragh 2007 pp 83-86).

267). However, these cosmological considerations had little impact on the physics community and ther is no evidence Einstein was aware of them.[6]

The end of the 19th century also saw a reconsideration of puzzles associated with Newtonian cosmology in the context of the new concepts of gravitational field and potential. Defining the gravitational potential $\Phi$ as

$$\Phi = G \int \frac{\rho(r)}{r} dV \qquad (1)$$

where G is Newton's gravitational constant and $\rho$ is the density of matter in a volume *V*, Newton's law of gravitation could be rewritten in terms of Poisson's equation

$$\nabla^2 \Phi = 4\pi G \rho \qquad (2)$$

where $\nabla^2$ is the Laplacian operator. Distinguished physicists such as Carl Neumann, Hugo von Seeliger and William Thomson noted that the gravitational potential would not be uniquely defined at all distances from a distribution of matter (Neumann 1896 pp 373-379; Seeliger 1985,1896; Thomson 1901). Neumann and Seeliger suggested independently that the problem could be solved by replacing Poisson's equation (2) with the relation

$$\nabla^2 \Phi - \lambda \Phi = 4\pi G \rho \qquad (3)$$

where $\lambda$ was a decay constant sufficiently small to make the modification significant only at extremely large distances.[7] A different solution to the problem was proposed in 1908 by the Swedish astronomer Carl Charlier, who considered a hierarchical or fractal structure for the universe; in this model the mean density of matter would tend to zero while the density would remain finite in every local location (Charlier 1908). This proposal was later taken up by Franz Selety, who argued that the hierarchic universe could provide a static, Newtonian cosmology alternate to Einstein's relativistic universe (Norton 1999).

*(iii) Relativistic cosmology and the problem of boundary conditions at infinity*

---

[6] See (Kragh 2012a,b) for a review of pre-1917 models of the universe of non-Euclidean geometry and their impact.
[7] See (North 1965 pp 17-18) or (Norton 1999) for a review of the Neumann-Seeliger proposal.

In 1915, Einstein published a set of covariant field equations that specified the relation between the geometry of a region of space-time and the distribution of matter/energy within it according to

$$G_{\mu\nu} = -\kappa \left( T_{\mu\nu} - \frac{1}{2} g_{\mu\nu} T \right) \quad (4)$$

where $G_{\mu\nu}$ is a four-dimensional tensor representing the curvature of space-time (known as the Ricci curvature tensor), $T_{\mu\nu}$ is the energy-momentum tensor, $T$ is a scalar and $\kappa$ is the Einstein constant $8\pi G/c^2$ (Einstein 1915). A description of Einstein's long path to his covariant field equations can be found in reviews such as (Norton 1984; Hoefer 1994; Janssen 2005; Janssen and Renn 2007). As noted in those references, Einstein's thoughts on Mach's principle and the relativity of inertia played a key role in the development of the theory. Indeed, in his well-known *'Prinzipielles'* paper of 1918, Einstein explicitly cited the principle as one of three principles[8] fundamental to the development of the theory: *"The G-field is completely determined by the masses of the bodies. Since mass and energy are – according to the results of the special theory of relativity – the same, and since energy is formally described by the symmetric energy tensor, it follows that the G-field is caused and determined by the energy tensor of matter"* (Einstein 1918a). Further insight into Einstein's understanding of Mach's principle and its relevance to cosmology is offered in the same article: *"Mach's Principle (c) is a different story. The necessity to uphold it is by no means shared by all colleagues: but I myself feel it is absolutely necessary to satisfy it. With (c), according to the field equations of gravitation, there can be no G-field without matter. Obviously postulate (c), is closely connected to the space-time structure of the world as a whole, because all masses in the universe will partake in the generation of the G-field"* (Einstein 1918a).

Even before the field equations had been published in their final, covariant form, Einstein had obtained an approximate solution for the case of the motion of the planets about the sun (Einstein 1915). In this calculation, the planetary orbits were modelled as motion around a point mass of central symmetry and it was assumed that at an infinite distance from that point, the metric tensor $g_{\mu\nu}$ would revert to flat 'Minkowski' space-time. Indeed, the orbits of the planets were calculated by means of a series of simple deviations from the Minkowski metric. The results corresponded almost exactly with the predictions of Newtonian mechanics with one exception; general relativity predicted an advance of 43" per century in the perihelion of the planet Mercury (Einstein 1915). This prediction marked the first success of the general theory, as the anomalous behaviour of Mercury had been well-known to astronomers for some years but had remained unexplained in Newton's theory. The result was a source of great satisfaction to Einstein and a strong indicator that his new theory of gravity was on the right track (Earman and Janssen 1993).

---

[8] The other principles cited were the principle of relativity and the principle of equivalence (Einstein 1918a).

In early 1916, Karl Schwarzschild obtained the first exact solution to the general field equations, again pertaining to the case of a mass point of central symmetry (Schwarzschild 1916). Einstein was surprised and delighted by the solution, declaring in a letter to Schwarzschild in January 1916 that *"I would not have expected that the exact solution to the problem could be formulated so simply"* (Einstein 1916b). In the Schwarzschild solution, it was once again assumed that sufficiently far from a material body, the space-time metric would revert to flat space-time. The imposition of such 'boundary conditions' was not unusual in field theory; however, such an approach could hardly be applied to the universe as a whole, as it raised the question of the existence a privileged frame of reference at infinity. Moreover, the assumption of a Minkowski metric an infinite distance away from matter was in obvious conflict with Einstein's understanding of Mach's principle.

Einstein's correspondence suggests that he continued to muse on the problem of boundary conditions at infinity throughout the year 1916. For example, a letter written to his old friend Michele Besso in May 1916 contains a reference to the problem, as well as an intriguing portend of Einstein's eventual solution: *"In gravitation, I am now looking for the boundary conditions at infinity; it certainly is interesting to consider to what extent a finite world exists, that is, a world of naturally measured finite extension in which all inertia is truly relative"* (Einstein 1916c). In the autumn of 1916, Einstein visited Leiden in Holland for a period of three weeks. There he spent many happy hours discussing his new theory of gravitation with his great friends Henrik Lorentz and Paul Ehrenfest. Also present at these meetings was the Dutch astronomer and theorist Willem de Sitter. A number of letters and papers written shortly afterwards by de Sitter (de Sitter 1916a,b) suggest that many of these discussions concerned the problem of boundary conditions, i.e., the difficulty of finding boundary conditions at infinity that were consistent with the principle of relativity and with Mach's principle. In one such article, de Sitter gives evidence that, at this stage, Einstein's solution was to suggest that, at an infinite distance from gravitational sources, the components of the metric tensor $[g_{\mu\nu}]$ would reduce to degenerate values: *"Einstein has, however, pointed* *out a set of degenerated $g_{ij}$ which are actually invariant for all transformations in which, at infinity $x_4$ is a pure function of $x'_4$. They are:*

$$\begin{pmatrix} 0 & 0 & 0 & \infty \\ 0 & 0 & 0 & \infty \\ 0 & 0 & 0 & \infty \\ \infty & \infty & \infty & \infty^2 \end{pmatrix}$$

*.... These are then the "natural " values, and any deviation from them must be due to material sources....At very large distances from all matter the $g_{ij}$ would gradually converge towards the degenerated values"* (de Sitter 1916a).

However, de Sitter highlights a potential flaw in Einstein's proposal. Since observation of the most distant stars showed no evidence of spatial curvature, it was puzzling how the 'local' Minkowskian values of the gravitational potentials $g_{\mu\nu}$ arose from the postulated degenerate values at infinity. According to de Sitter, Einstein proposed that this effect was due to the influence of distant masses: *"Now it is certain that, in many systems of reference (i.e., in all Galilean systems) the $g_{ij}$ at large distances from all material bodies known to us actually have the [Minkowski] values. On Einstein's hypothesis, these are special values which, since they differ from [degenerate] values, must be produced by some material bodies. Consequently there must exist, at still larger distances, certain unknown masses which are the source of the [Minkowski] values, i.e., of all inertia* (de Sitter 1916a). Yet no trace of such masses were observable by astronomy: *"We must insist on the impossibility that any of the known fixed stars or nebulae can form part of these hypothetical masses. The light even from the farthest stars and nebulae has approximately the same wavelength as light produced by terrestrial sources. ...the deviation of the $g_{ij}$ from the Galilean values ... is of the same order as here, and they must therefore be still inside the limiting envelope which separates our universe from the outer parts of space, where the $g_{ij}$ have the [degenerate] values".* Indeed, de Sitter concludes that the hypothetical distant masses essentially play the role of absolute space in classical theory. *"If we believe in the existence of these supernatural masses, which control the whole physical universe without having ever being observed then the temptation must be very great indeed to give preference to a system of co-ordinates relatively to which they are at rest, and to distinguish it by a special name, such as "inertial system" or "ether". Formally the principle of relativity would remain true, but as a matter of fact we would have returned to the absolute space under another name"* (de Sitter 1916a).

Einstein and de Sitter debated the issue of boundary conditions at infinity in corrsepondence for some months. A review of their fascinating debate can be found in references such as (Kerzberg 1989; Hoefer 1994; Schulmann et al. 1998 pp 353-354; Realdi and Peruzzi 2009). We note here that Einstein conceded defeat on the issue in a letter written to de Sitter on November 4[th] 1916: *"I am sorry for having placed too much emphasis on the boundary conditions in our discussions. This is purely a matter of taste which will never gain scientific significance. ……Now that the covariant field equations have been found, no motive remains to place such great weight on the total relativity of inertia* (Einstein 1916d). However, the closing paragraph of the same letter indicates that Einstein had not completely given up on the notion of the relativity of inertia: *"On the other hand, you must not scold me for being curious enough still to ask: Can I imagine a universe or the universe in such a way that inertia stems entirely from the masses and not at all from the boundary conditions? As long as I am aware that this whim does not touch the core of the theory, it is innocent; by no means do I expect you to share this curiosity"* (Einstein 1916d). Notice of a successful conclusion to Einstein's quest appears in another letter to de Sitter, written on 2[nd] February 1917: "*Presently I am writing a paper on the boundary conditions in gravitation theory. I have completely abandoned my idea on the degeneration of the $g_{\mu\nu}$, which you rightly disputed. I am curious to see what you will say about the rather outlandish conception I have now set my sights on*" (Einstein 1917c). The 'outlandish conception' was the postulate of a universe of closed spatial geometry, as described below.

## 3. Einstein's 1917 paper

A surprising feature of Einstein's 1917 cosmological memoir is the sizeable portion of the paper concerned with Newtonian cosmology. This analysis had two important aims. In the first instance, Einstein was no doubt pleased to show that his new theory of gravitation could overcome a well-known puzzle associated with Newtonian cosmology. Second, a suggested *ad-hoc* modification of Newtonian gravity provided a useful analogy for a necessary modification of the field equations of relativity.

Einstein's assault on Newtonian cosmology is two-pronged. First he establishes from symmetry principles that Newtonian gravity only allows for a finite island of stars in infinite space. Then he suggests from a consideration of statistical mechanics that such an island would evaporate, in contradiction with the presumed static nature of the universe. His solution to the paradox is the introduction of a new term to Poisson's equation. This solution is very similar to that of Seeliger and Neumann, although Einstein was not aware of this work at the time (O'Raifeartaigh et al. 2017). A year later, Einstein presented a simpler argument against the Newtonian universe in terms of lines of force; this argument was published in the third edition of his popular book on relativity (Einstein 1918b p123) and retained in all later editions of the book.

We note that a few years after the publication of the 1917 memoir, the Austrian physicist Franz Selety noted that the hierarchic cosmology proposed by Carl Charlier (above) avoided the paradox identified by Einstein (Selety 1922). Einstein conceded the point, but objected to the Charlier's model on the grounds that it was anti-Machian (Einstein 1922b).[9]

*(i)     On the basic assumptions of Einstein's model*

It is clear from Einstein's 1917 memoir that the starting point of his cosmic model was the assumption of a universe with a static distribution of matter, uniformly distributed over the largest scales and of non-zero average density. Considering the issue of stasis first, Einstein argued for a quasi-static distribution of matter based on the small velocities of the stars: *"The most important fact that we draw from experience as to the distribution of matter is that the relative velocities of the stars are very small as compared with the velocity of light. So ....there is a system of reference relatively to which matter may be looked upon as being permanently at rest"* (Einstein 1917a). It is generally agreed amongst historians and physicists that this assumption was reasonable at the time (Hoefer 1994; Kragh 2007 pp 131-132; Nussbaumer and Bieri 2009 pp 72-76). There is no evidence that Einstein was aware at this time of Slipher's observations of the redshift of light from the spiral nebulae, while the extra-galactic nature of the spirals had yet to be established. Indeed, many years were to elapse before the demonstration of a linear relation between the redshifts of the distant galaxies and their distance (Hubble 1929), the first evidence for a non-static universe. However, it's worth noting that Einstein's stellar argument was questioned by de Sitter: *"We only have a snapshot of the world, and we cannot and must not conclude from the fact that we do not see any large changes on this photograph that everything will always remain as at that instant when the picture was taken."* (de Sitter 1917b). It could also be argued that Einstein erred philosophically in inferring *global* stasis from astronomical observations of the *local* environment (Kerzberg 1989; Smeenk 2014 p241); however, we find his assumption reasonable in the context of the widespread contemporaneous belief that the universe was not much larger than the Milky Way.

---

[9] See (Norton 1999) for a discussion of the Einstein-Selety debate.

It is sometimes stated that Einstein's assumption of stasis prevented him from predicting the expansion of the universe many years before the phenomenon was discovered by astronomers. This statement may be true in a literal sense, but we find it somewhat anachronistic. It is clear throughout his cosmological memoir that Einstein's interest lay in establishing whether he could achieve a description of the universe consistent with the foundational principles of general relativity (including, in particular, Mach's principle) and with astronomical observation. Thus, the exploration of solutions to the field equations for the case of a non-static cosmos would have been of little interest to him in 1917. Many years later, Einstein stated that the assumption of a static universe *"appeared unavoidable to me at the time, since I thought that one would get into bottomless speculations if one departed from it"* (Einstein 1945 p137). Indeed, it could be argued that the common moniker 'Einstein's static model of the universe' is a little misleading, as it implies a choice from a smorgasbord of possible models of the known universe. Historically speaking, a more accurate title would be 'Einstein's model of the Static Universe'.

In some ways, Einstein's assumption of matter "*as being uniformly distributed over enormous spaces*" was more radical than his assumption of stasis. Technically speaking, this assumption implied a universe that was both isotropic and homogeneous, at least on the largest scales, an assumption that was at odds with astronomical observations. Thus, the assumption was more of an assumed principle and indeed it was later named the 'Cosmological Principle' (Milne 1935 p24). One reason for the principle was its undoubted simplicity, as the assumption of homogeneity and isotropy greatly simplified the business of solving the field equations. A deeper reason may have been that the Cosmological Principle chimed with a Copernican approach to cosmology and with the spirit of relativity (Bondi 1952 pp 11-13). After all, to assume a universe with a non-uniform distribution of matter on the largest scales was to assume a universe in which all viewpoints were not equivalent, in contradiction with basic tenets of relativity (Milne 1933).

*(ii)   On spatial curvature*

In his 1917 memoir, Einstein's solution to the problem of boundary conditions at infinity was to banish the boundaries by postulating a world of closed, spherical spatial curvature. In this manner, the Einstein World explicitly incorporated his view of the relativity of inertia.[10] It was later shown that closed geometry was the *only* possibility for a universe with a static, homogeneous distribution of matter of non-zero average density. Thus, Einstein's view of Mach's principle was a useful, but not strictly necessary, guide to his first model of the universe, just as it was a guide on his path to the field equations.

---

[10] Unfortunately, Einstein's expression *"räumlich geschlossen"* or "spatially closed" is mistranslated throughout the official English translation of the paper as "spatially finite" (Einstein 1917a; O'Raifeartaigh et al. 2017).

Following the publication of the 1917 paper, colleagues such as Erwin Freundlich, Felix Klein and Willem de Sitter suggested in correspondence to Einstein that elliptical geometry would also satisfy the requirements of his cosmology (O'Raifeartaigh et al. 2017). Einstein quickly conceded the point, noting that the relation between the radius of curvature and the mean density of matter remained unchanged. For example, he remarked to Klein: *"As I have never done non-Euclidean geometry, the more obvious elliptical geometry had escaped me….my observations are just altered thus, that the space is half as large; the relation between R (the radius of curvature) and ρ (mean density of matter) is retained"* (Einstein 1917d). A few months later, he commented to de Sitter: *"When I was writing the paper, I did not yet know about the elliptical possibility…..this possibility seems more likely to me as well"* (Einstein 1917e). This preference was cited by de Sitter in his classic paper of 1917: *"The elliptical space is, however, really the simpler case, and it is preferable to adopt this for the physical world"* …*this is also the opinion of Einstein"* (de Sitter 1917a). Neither Einstein nor de Sitter make clear in their correspondence why they prefer elliptical geometry; one explanation may be that they viewed this geometry as more general than spherical.

*(iii)   On the cosmological constant*

In his cosmological memoir, Einstein soon found that the hypothesis of closed spatial geometry was not sufficient to achieve a successful relativistic model of the universe. Instead, a consistent solution could only be achieved with the introduction of an additional term $\lambda g_{\mu\nu}$ to the field equations, according to

$$G_{\mu\nu} - \lambda g_{\mu\nu} = -\kappa \left( T_{\mu\nu} - \frac{1}{2} g_{\mu\nu} T \right)$$

(5)

where λ was a universal constant that became known as the *cosmological constant*. Einstein then showed that the modified field equations (5) have the solution

$$\lambda = \frac{\kappa \rho}{2} = \frac{1}{R^2} \qquad (6)$$

where $\rho$ and $R$ represent the mean density of matter and the radius of the cosmos respectively (Einstein 1917a). In this manner, Einstein's 1917 model of the cosmos gave an apparently satisfactory relation between the size of the universe and the amount of matter it contained.

Thus, Einstein's model appears to have evolved according to the following sequence of assumptions: uniform, static distribution of matter → closed spatial geometry → introduction of additional term to the field equations. While the general theory allowed such a modification of the field equations, Einstein seems to have anticipated some resistance to the term; it is interesting that he forewarns the reader of what is to come on three separate occasions in the paper. Indeed, it could be argued that much of Einstein's 1917 memoir can be read as a lengthy justification for the introduction of the cosmological constant term to relativity!

Some historians have found Einstein's use of the cosmological constant term in his 1917 memoir somewhat ambiguous and argue that his view of the term wavers throughout the paper (Kerzberg 1989). In our view, the *purpose* of the term is clear throughout the paper, both in the stated text and in the underlying physics of the model, and is summarized quite precisely in the final sentence: *"That term is necessary only for the purpose of making possible a quasi-static distribution of matter, as required by the fact of the small velocities of the stars"*. That said, there is little doubt that the cosmological constant term posed a significant challenge to Einstein in terms of interpretation. Indeed, it is striking that no interpretation of the physics underlying the term is presented anywhere in the 1917 paper and there is ample evidence in Einstein's later writings that he viewed his modification of the field equations as an uncomfortable mathematical necessity. For example, in March 1917, Einstein remarked to Felix Klein: *"The new version of the theory means, formally, a complication of the foundations and will probably be looked upon by almost all our colleagues as an interesting, though mischievous and superfluous stunt, particularly since it is unlikely that empirical support will be obtainable in the foreseeable future. But I see the matter as a necessary addition, without which neither inertia nor geometry are truly relative"* (Einstein 1917d). Similarly, when de Sitter commented in a letter of March 20[th]: " *I personally much prefer the four-dimensional system, but even more so the original theory, without the undeterminable λ, which is just philosophically and physically desirable* (de Sitter 1917c), Einstein responded: *"In any case, one thing stands. The general theory of relativity allows the addition of the term $\lambda g_{\mu\nu}$ in the field equations. One day, our actual knowledge of the composition of the fixed-star sky, the apparent motions of* *fixed stars, and the position of spectral lines as a function of distance, will probably have come far enough for us to be able to decide empirically the question of whether or not λ vanishes. Conviction is a good mainspring, but a bad judge!"* (Einstein 1917f).

In March 1918, the Austrian physicist Erwin Schrödinger suggested that a consistent model of a static, matter-filled cosmos could be obtained from Einstein's field equations without the introduction of the cosmological constant term (Schrödinger 1918). Essentially, Schrödinger's proposal was that Einstein's solution could be obtained from the unmodified field equations (4) if a negative-pressure term was added to the 'source' tensor on the right-hand side of the equations, i.e., by replacing Einstein's energy-momentum tensor by the tensor

$$T^{\mu\nu} = \begin{pmatrix} -p & 0 & 0 & 0 \\ 0 & -p & 0 & 0 \\ 0 & 0 & -p & 0 \\ 0 & 0 & 0 & \rho - p \end{pmatrix} \quad (7)$$

where $\rho$ is the mean density of matter and $p$ is the pressure (defined as $p = \lambda/\kappa$ ).

Einstein's response was that Schrödinger's formulation was entirely equivalent to that of his 1917 memoir, provided the negative-pressure term was constant (Einstein 1918c).[11] This response seems at first surprising; Schrödinger's new term may have been *mathematically* equivalent to that of Einstein's but the underlying physics was surely different. However, in the same paper, Einstein gave his first physical interpretation of the cosmological term, namely that of a negative mass density: *"In terms of the Newtonian theory…a modification of the theory is required such that "empty space" takes the role of gravitating negative masses which are distributed all over the interstellar space"* (Einstein 1918c).

Within a year, Einstein proposed a slightly different interpretation of the cosmological constant term. Rewriting the field equations in a slightly different format, he opined that the cosmological constant now took the form of a constant of integration, rather than a universal constant associated with cosmology: *"But the new formulation has this great advantage, that the quantity appears in the fundamental equations as a constant of integration, and no longer as a universal constant peculiar to the fundamental law"* (Einstein 1919). Indeed, a letter to Michele Besso suggests that Einstein had arrived at a similar interpretation a year earlier using a variational principle (Einstein 1918d). A follow-up letter to Besso suggests that at one point, Einstein considered the two views to be equivalent: *"Since the world exists as a single*

---

[11] Schrödinger also suggested that the pressure term might be time variant, anticipating the modern concept of quintessence, but this suggestion was too speculative for Einstein (Einstein 1918c).

*specimen, it is essentially the same whether a constant is given the form of one belonging to the natural laws or the form of an 'integration constant'"* (Einstein 1918e).

Thus, there is little doubt that a satisfactory interpretation of the physics of the cosmological constant term posed a challenge for Einstein in these years. One contributing factor to this ambiguity may be a slight mathematical confusion concerning manner in which the term was introduced. As several scholars have noted (Norton 1999; Harvey and Schucking 2000), Einstein's modification of the field equations in his memoir was not in fact exactly analogous to his modification of Newtonian gravity, as he claimed, i.e., the modified field equations (5) do not reduce in the Newtonian limit to the modified Poisson equation (3), but to the slightly different relation

$$\nabla^2 \phi + c^2 \lambda = 4\pi G \rho \qquad (8)$$

This might seem a rather pedantic point, but the error may have been significant with regard to Einstein's interpretation of the term. Where he intended to introduce a term to the field equations representing an attenuation of the gravitational interaction at large distances, he in fact introduced a term representing a tendency for empty space to expand, a concept that would have been in conflict with his view of Mach's principle at the time.

*(iv)    On testing the model against observation*

A curious aspect of Einstein's 1917 memoir is that, having established a pleasing relation between the geometry of the universe and the matter it contained, he made no attempt to test the model against empirical observation. After all, even a rough estimate of the mean density of matter $\rho$ in equation (6) would give a value for the cosmic radius $R$ and the cosmological constant $\lambda$. These values could then have been checked against observation; one could expect an estimate for $R$ that was not smaller than astronomical estimates of the size of the distance to the furthest stars, and an estimate for $\lambda$ that was not too large to be compatible with observations of the orbits of the planets. No such calculation is to be found in the 1917 memoir. Instead, Einstein merely declares at the end of the paper that the model is logically consistent: *"At any rate, this view is logically consistent, and from the standpoint of the general theory of relativity lies nearest at hand; whether, from the standpoint of present astronomical knowledge, it is tenable, will not here be discussed"*.

We have previously noted that Einstein did attempt such a calculation in his correspondence around this time (O'Raifeartaigh et al. 2017). Taking a value of $\rho = 10^{-22}$ g/cm$^3$ for the mean density of matter,[12] he obtained from equation (6) an estimate of $10^7$ light-years for the radius of his universe, a result he found unrealistic. As he stated in a letter to Paul Ehrenfest: *"From the measured stellar densities, a universe radius of the order of magnitude of $10^7$ light-years results, thus unfortunately being very large against the distances of observable stars"* (Einstein 1917g). This comment implies that, like many of his contemporaries at the time, Einstein did not believe that the universe was significantly larger than the Milky Way. However, Einstein does not appear to have taken such calculations too seriously, presumably because he lacked confidence in astronomical estimates of the mean density of matter. As he remarked in a letter to Erwin Freundlich: *"..The matter of great interest here is that not only R but also ρ must be individually determinable astronomically, the latter quantity at least to a very rough approximation, and then my relation between them ought to hold. Maybe the chasm between the $10^4$ and $10^7$ light years can be bridged after all. That would mean the beginning of an epoch in astronomy"* (Einstein 1917h). Later writings also suggest that Einstein viewed the average density of matter in the universe as an unknown quantity (Einstein 1921; O'Raifeartaigh et al. 2017).

*(v)     On the stability of the Einstein World*

Perhaps the strangest aspect of Einstein's 1917 memoir is his failure to consider the stability of his cosmic model. After all, equation (6) drew a direct equation between a universal constant *λ*, the radius of the universe *R*, and the density of matter *ρ*. But the quantity *ρ* represented a *mean* value for the density of matter, arising from the theoretical assumption of a uniform distribution of matter on the largest scales. In the real universe, one would expect a natural variation in this parameter in time and space, raising the question of the stability of the model against such perturbations. It was later shown that the Einstein World is generally unstable against such density perturbations: instead of oscillating around a stable solution, a slight increase in the density of matter (without a corresponding change in *λ*) would cause the universe to contract, become more dense and contract further, while a slight decrease in density would result in a runaway expansion (Eddington 1930).[13]

---

[12] Einstein does not give a reference for his estimate of the mean density of matter in his correspondence but it is in reasonable agreement with that given by de Sitter (de Sitter 1917a).
[13] See (Gibbons 1987) for further discussion of the stability of the Einstein World.

It is curious that Einstein did not consider this aspect of his model in 1917; some years later, it was a major reason for his rejection of the model, as described in the next section.

## 4. The Einstein – de Sitter debate

In July 1917, Willem de Sitter published a paper in which he noted that the modified field equations allowed a cosmological solution for the case of a universe with no matter content (de Sitter 1917a). In this cosmology, Einstein's matter-filled three-dimensional universe of spherical spatial geometry was replaced by an empty four-dimensional universe of closed *spacetime* geometry. It should come as no surprise that Einstein was greatly perturbed by de Sitter's solution, as the model was in direct conflict with his understanding of Mach's principle in these years. A long debate ensued between the two physicists concerning the relative merits of the two models that has been extensively described in the literature.[14] Eventually, Einstein made his criticisms public in a paper of 1918: *"It appears to me that one can raise a grave argument against the admissibility of this solution…..In my opinion, the general theory of relativity is a satisfying system only if it shows that the physical qualities of space are completely determined by matter alone. Therefore no $g_{\mu\nu}$- field must exist (that is no space-time continuum is possible) without matter that generates it"* (Einstein 1918f). Einstein also raised a technical objection to de Sitter's model, namely that it appeared to contain a spacetime singularity. In the years that followed, Einstein continued to debate the de Sitter model with physicists such as Kornel Lanczos, Hermann Weyl, Felix Klein and Gustav Mie. Throughout this debate, Einstein did not waver from his core belief that a satisfactory cosmology should describe a universe that was globally static with a metric structure that was fully determined by matter.[15] Einstein eventually conceded that the apparent singularity in the de Sitter universe was an artefact of co-ordinate representation (Einstein 1918g), but he never formally retracted his criticism of the de Sitter universe in the literature, nor did he refer to the de Sitter model in his formal writings on cosmology in these years (O'Raifeartaigh et al. 2017).

## 5. Einstein and the expanding universe

---

[14] See for example (Kerzberg 1989; Schulmann et al. 1998 pp 352-354; Realdi and Peruzzi 2009).
[15] See (Schulmann et al. 1988 pp 355-357) for a discussion of the Einstein-deSitter-Weyl-Klein debate.

In 1922, the Russian physicist Alexander Friedman suggested that non-static solutions of the Einstein field equations should be considered in relativistic models of the cosmos (Friedman 1922). Einstein publicly faulted Friedman's analysis on the basis that it contained a mathematical error (Einstein 1922a). When it transpired that the error lay in Einstein's criticism, it was duly retracted (Einstein 1923a). However, an unpublished draft of Einstein's retraction demonstrates that he did not consider Friedman's cosmology to be realistic: *"to this a physical significance can hardly be ascribed"* (Einstein 1923b).[16]

A few years later, the Belgian physicist Georges Lemaître independently derived time-varying equations for the radius of the cosmos from Einstein's modified field equations. Aware of Slipher's observations of the redshifts of the spiral nebulae, and of emerging measurements of the distance of the spirals by Edwin Hubble, Lemaître suggested that the recession of the nebulae was a manifestation of the expansion of space from a pre-existing Einstein World of cosmic radius $R_0 = 1/\sqrt{\lambda}$ (Lemaître 1927). This work was brought to Einstein's attention by Lemaître himself, only to have expanding cosmologies dismissed as *"abominable"*. According to Lemaître, Einstein's rejection probably stemmed from a lack of knowledge of developments in astronomy: *"Je parlais de vitesses des nébeleuses et j'eus l'impression que Einstein n'était guère au courant des faits astronomiques"* (Lemaître 1958).

In 1929, Edwin Hubble published empirical evidence of a linear relation between the redshifts of the spiral nebulae and their radial distance (Hubble 1929).[17] Many theorists interpreted the observations in terms of a relativistic expansion of space, and a number of cosmic models of the Friedman-Lemaître type were advanced for diverse values of cosmic parameters. Einstein himself overcame his earlier distrust of expanding models of the cosmos, stating during a sojourn at the California Institute of Technology in 1931: *"New observations by Hubble and Humason concerning the redshift of light in distant nebulae make the presumptions near that the general structure of the universe is not static"* (AP 1931a) and *"The redshift of the distant nebulae have smashed my old construction like a hammer blow"* (AP 1931b). A recently-discovered manuscript indicates that Einstein first considered a steady-state model of the universe on learning of Hubble's observations; however the model led to a null solution and he quickly abandoned the attempt (O'Raifeartaigh et al. 2014; Nussbaumer 2014a). In April 1931, Einstein published a model of the expanding cosmos based on Friedman's 1922 analysis, with the cosmological term removed, deriving simple expressions relating the rate of cosmic expansion (an observable that could be measured from the recession of the nebulae) to the radius of the cosmos, the density of matter and the timespan of the expansion.[18] It is interesting to note that Einstein provided a two-fold justification for abandoning the cosmological constant term in this paper. In the first instance, the term was unsatisfactory because it did not provide a stable solution: *"It can also be shown… that this solution is not stable. On these grounds alone, I am no longer inclined to ascribe a physical meaning to my former solution"* (Einstein 1931). In the second instance, the term was unnecessary because the assumption of stasis was no longer justified by observation: *"Now that it has become clear from Hubbel's [sic] results that the extra-galactic nebulae are uniformly distributed throughout space and are in dilatory motion (at least if their systematic redshifts are to be interpreted as Doppler effects), assumption (2) concerning the static nature of space has no longer any justification"* (Einstein 1931). A year later, Einstein proposed an even simpler model of the expanding universe in conjunction with de Sitter; in this model, both the cosmological constant and spatial curvature were removed (Einstein and de Sitter 1932).

---

[16] A detailed account of this episode can be found in (Nussbaumer and Bieri 2009 pp 91-92).
[17] Although Lemaître had derived such a relation in 1927 from theory, the empirical verification of the relation is attributable to Hubble (O'Raifeartaigh 2014; Kragh 2018).

Thus it is clear that, when presented with empirical evidence for a dynamic universe, Einstein lost little time in abandoning his static cosmology.[19] He also abandoned the cosmological constant term and was never to re-instate it in his cosmological models. Indeed, he is reputed to have described the term in later years as his *"biggest blunder"*. Whether Einstein used these exact words has been the subject of some debate,[20] but his considered view of the cosmological constant term was made clear in a 1945 review of relativistic cosmology:*"If Hubble's expansion had been discovered at the time of the creation of the general theory of relativity, the cosmologic member would never have been introduced. It seems now so much less justified to introduce such a member into the field equations, since its introduction loses its sole original justification – that of leading to a natural solution of the cosmologic problem"* (Einstein 1945 p130). This passage neatly encapsulates Einstein's matter-of-fact approach to cosmology - if the known universe could be modelled without the cosmological constant term, why include it?

## 6.    Conclusions

---

[18] We have recently provided an analysis and first English translation of this paper (O'Raifeartaigh and McCann 2014) and noted that Einstein's calculations contain a systematic error.
[19] See (Nussbaumer 2014b) for further details on Einstein's conversion to expanding cosmologies.
[20] We have recently provided an interrogation of this story (O'Raifeartaigh and Mitton 2018).

In his 1917 cosmological memoir, Einstein demonstrated that his newly-minted general theory of relativity could give a model of the universe that was consistent with the founding principles of the theory, including Mach's principle, and with astronomical observation. The price was the hypothesis of closed spatial geometry for the cosmos and a modification of the field equations of general relativity. A slight mathematical inaccuracy associated with Einstein's introduction of the cosmological constant is intriguing; it is possible that this ambiguity may have affected his interpretation of the term. It is also interesting that Einstein made no formal attempt to test his model against empirical observation; later writings suggest that he distrusted astronomical estimates of the mean density of matter in the universe. Perhaps the most curious aspect of Einstein's 1917 memoir is his failure to consider the stability of his cosmic model. When he formally abandoned the Einstein World in 1931, it was on the twin grounds that the model was both theoretically unstable and in conflict with empirical observation.

We note finally that the Einstein World has become a topic of renewed interest in today's cosmology. Some theorists have become interested in the hypothesis of a universe that expands from a static Einstein World after an indefinite period of time, thus reviving Lemaître's 1927 model in the context of the modern theory of cosmic inflation. It is thought that this scenario, known as 'the emergent universe', might be useful in addressing major difficulties in modern cosmology such as the horizon problem, the quantum gravity era and the initial singularity.[21] Whether the emergent universe will offer a plausible, consistent description of the origins and evolution of our universe is not yet known, but we note, as so often, the relevance of past models of the universe in today's research.

**Acknowledgements**

The author acknowledges the use of The Collected Papers of Albert Einstein (CPAE), an invaluable online archive of Einstein's original works complete with annotations and editorial comments, provided by Princeton University Press in conjunction with the California Institute of Technology and the Hebrew University of Jerusalem. Helpful comments from the editors and an anonymous referee are also acknowledged.

---

[21] See (Barrow et al. 2003) for an introduction to emergent cosmology.